\documentclass[english,journal,final,twocolumn,10pt,twoside]{IEEEtran}
\usepackage[T1]{fontenc}
\usepackage[latin9]{inputenc}
\usepackage{amsmath}
\usepackage{graphicx}

\makeatletter
\adddialect\l@ENGLISH\l@english
\usepackage{url}

\makeatother

\usepackage{babel}
\begin{document}
\markboth{IEEE Signal Processing Letters, Vol. 14, No. 12, December 2007}{Valin and Collings: Interference-Normalised Least Mean Square Algorithm}

\title{Interference-Normalised Least Mean Square Algorithm}

\author{Jean-Marc Valin\emph{, \IEEEmembership{Member, IEEE},} and Iain
B. Collings\inputencoding{latin1}{, }\inputencoding{latin9}\IEEEmembership{Senior Member, IEEE}\thanks{The authors are with CSIRO ICT Centre, Marsfield, NSW, 2122, Australia (e-mail: jmvalin@jmvalin.ca; iain.collings@csiro.au).}\thanks{\copyright 2007 IEEE.  Personal use of this material is permitted. Permission from IEEE must be obtained for all other uses, in any current or future media, including reprinting/republishing this material for advertising or promotional purposes, creating new collective works, for resale or redistribution to servers or lists, or reuse of any copyrighted component of this work in other works.}}
\maketitle
\begin{abstract}
\setcounter{page}{988}An interference-normalised least mean square
(INLMS) algorithm for robust adaptive filtering is proposed. The INLMS
algorithm extends the gradient-adaptive learning rate approach to
the case where the signals are non-stationary. In particular, we show
that the INLMS algorithm can work even for highly non-stationary interference
signals, where previous gradient-adaptive learning rate algorithms
fail. \end{abstract}

\begin{keywords}
NLMS, gradient-adaptive learning rate, adaptive filtering

\begin{center} \bfseries EDICS Category: SAS-ADAP \end{center}

\IEEEpeerreviewmaketitle
\end{keywords}

\section{Introduction}

The choice of learning rate is one of the most important aspects of
least mean square adaptive filtering algorithms as it controls the
trade off between convergence speed and divergence in presence of
interference. 

In this paper, we introduce a new interference-normalised least mean
square (INLMS) algorithm. In the same way as the NLMS algorithm introduces
normalisation against the filter input $x\left(n\right)$, our proposed
INLMS algorithm extends the normalisation to the interference signal
$v\left(n\right)$. The approach is based on the gradient-adaptive
learning rate class of algorithms \cite{Benveniste,Mathews1993,Ang2001,Mandic2004},
but improves upon these algorithms by being robust to non-stationary
signals.

\begin{figure}
\begin{center}\includegraphics[width=65mm,keepaspectratio]{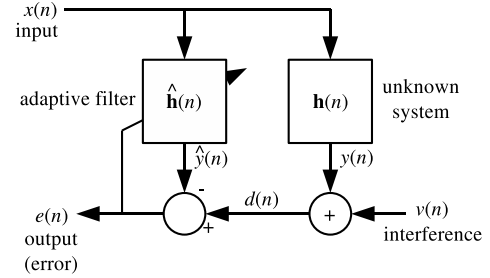}\end{center}

\caption{Block diagram of echo cancellation system.\label{cap:Block-diagram}}
\end{figure}

We consider the adaptive filter illustrated in Fig. \ref{cap:Block-diagram},
where the input signal $x\left(n\right)$ is convolved by an unknown
$\mathbf{h}\left(n\right)$ filter (to produce $y\left(n\right)$)
which has an additive interference signal signal $v\left(n\right)$,
before being observed as $d\left(n\right)$. The adaptive filter attempts
to estimate the impulse response $\mathbf{\hat{h}}\left(n\right)$
to be as close as possible to the real impulse response $\mathbf{h}\left(n\right)$
based only on the observable signals $x\left(n\right)$ and $d\left(n\right)$.
The estimated convolved signal $\hat{y}\left(n\right)$ is subtracted
from $d\left(n\right)$, giving an output signal $e\left(n\right)$
containing both the interference $v\left(n\right)$ and a residual
signal $r\left(n\right)=y\left(n\right)-\hat{y}\left(n\right)$. In
many scenarios, such as echo cancellation, the interference $v\left(n\right)$
is actually the signal of interest in the system. 

The standard normalised least mean squares (NLMS) algorithm is given
by:
\begin{align}
e\left(n\right) & =d\left(n\right)-\hat{\mathbf{h}}^{H}\left(n{-}1\right)\mathbf{x}\left(n\right)\\
\hat{\mathbf{h}}\left(n\right) & =\hat{\mathbf{h}}\left(n{-}1\right)+\frac{\mu\left(n\right)}{\left\Vert \mathbf{x}\left(n\right)\right\Vert ^{2}}e^{*}\left(n\right)\mathbf{x}\left(n\right)
\end{align}
where $\mathbf{x}\left(n\right)=\left[x\left(n\right),x\left(n-1\right),\dots,x\left(n-L+1\right)\right]^{T}$
and $\mu\left(n\right)$ is the learning rate. Here, we propose to
extend this algorithm, by adaptively updating $\mu\left(n\right)$.
By adopting our approach, we develop an algorithm which we call the
INLMS algorithm and which works even for highly non-stationary interference
signals, where previous gradient-adaptive learning rate algorithms
fail.

Section \ref{sec:Gradient-Adaptive-Learning-Rate} introduces existing
gradient-adaptive learning rate algorithms and their limitations.
Section \ref{sec:INLMS-algorithm} describes our proposed INLMS algorithm,
followed by the results and discussion in Section \ref{sec:Results-And-Discussion}.
Section \ref{sec:Conclusion} concludes this paper.

\section{Gradient-Adaptive Learning Rate\label{sec:Gradient-Adaptive-Learning-Rate}}

Gradient-adaptive learning rate algorithms are based on the fact that
when the adaptation rate is too small, the gradient tends to keep
pointing in the same direction, while if it is too large, the gradient
oscillates. Based on the behaviour of the stochastic gradient, it
is thus possible to infer whether the learning rate must be increased
or decreased, and several methods have been proposed in the past to
adjust the learning based on the gradient. 

These methods each have a \emph{control parameter} that is used to
determine the learning rate. In the case of \cite{Benveniste,Mathews1993,Ang2001},
the control parameter is the learning rate itself. In the generalized
normalized gradient descent (GNGD) algorithm \cite{Mandic2004} the
(normalised) learning rate is:
\begin{equation}
\mu\left(n\right)=\frac{\mu_{0}\left\Vert \mathbf{x}\left(n\right)\right\Vert ^{2}}{\left\Vert \mathbf{x}\left(n\right)\right\Vert ^{2}+\epsilon\left(n\right)}\label{eq:GNGD}
\end{equation}
where $\epsilon\left(n\right)$ is the control parameter. Because
the control parameter is adapted based on the NLMS stochastic gradient
behaviour, it can only vary relatively slowly (typically requiring
tens or hundreds of samples). For that reason, it is important for
the optimal learning rate not to depend on rapid changes of the control
parameter. We will show in the next section that none of the methods
cited above can fulfil this condition for non-stationary sources.

\subsection{Analysis For Non-Stationary Signals}

Under the assumption that $x\left(n\right)$ and $v\left(n\right)$
are zero-mean and uncorrelated to each other and that $v\left(n\right)$
is i.i.d., the theoretical optimal learning rate is equal to the residual-to-error
ratio \cite{ValinAEC2006}:
\begin{equation}
\mu_{opt}\left(n\right)=\frac{E\left\{ r^{2}\left(n\right)\right\} }{E\left\{ e^{2}\left(n\right)\right\} }\label{eq:Residual-to-output-rate}
\end{equation}
where $r\left(n\right)=y\left(n\right)-\hat{y}\left(n\right)$ is
the (unknown) residual echo and $e\left(n\right)$ is the error signal.
It turns out that although the assumption on $v\left(n\right)$ is
not verified for speech, (\ref{eq:Residual-to-output-rate}) nonetheless
remains a good approximation. Earlier gradient-adaptive algorithms
vary $\mu\left(n\right)$ directly as a response to the behaviour
of the gradient ($\mu\left(n\right)$ is the control parameter). It
is a sensible thing to do if one assumes that $v\left(n\right)$ and
$x\left(n\right)$ are stationary signals, because it means that both
$E\left\{ r^{2}\left(n\right)\right\} $ and $E\left\{ e^{2}\left(n\right)\right\} $
vary slowly and, as a consequence, so does $\mu_{opt}\left(n\right)$.
On the other hand, if the statistics of either $v\left(n\right)$
or $x\left(n\right)$ changes abruptly, then the algorithm is not
capable of changing $\mu\left(n\right)$ fast enough to prevent the
adaptive filter from diverging.

The GNGD algorithm provides more robustness to non-stationarity. If
we examine $\epsilon\left(n\right)$ more closely, it is reasonable
to surmise that (\ref{eq:GNGD}) eventually converges to the optimal
learning rate defined by (\ref{eq:Residual-to-output-rate}). Assuming
steady state behaviour ($\epsilon\left(n\right)$ is stable) and $\mu_{0}=1$,
we find (by multiplying the left hand side numerator and denominator
by $\gamma\left(n\right)$) that:
\begin{equation}
\frac{E\left\{ r^{2}\left(n\right)\right\} }{E\left\{ r^{2}\left(n\right)\right\} +\gamma\left(n\right)\epsilon\left(n\right)}=\frac{E\left\{ r^{2}\left(n\right)\right\} }{E\left\{ e^{2}\left(n\right)\right\} }
\end{equation}
where $\gamma\left(n\right)=E\left\{ r^{2}\left(n\right)\right\} /\left\Vert \mathbf{x}\left(n\right)\right\Vert ^{2}$
is analogous to the filter misalignment. Assuming that $r\left(n\right)$
and $v\left(n\right)$ are zero-mean and uncorrelated to each other,
we have $E\left\{ e^{2}\left(n\right)\right\} =E\left\{ r^{2}\left(n\right)\right\} +E\left\{ v^{2}\left(n\right)\right\} $,
which results in the relation $\epsilon\left(n\right)=E\left\{ v^{2}\left(n\right)\right\} /\gamma\left(n\right)$.
In other words, the optimal value for the gradient-adaptive parameter
$\epsilon\left(n\right)$ depends on the filter misalignment and on
the variance of the interference signal, but is independent of the
variance of the input signal. Because $\epsilon\left(n\right)$ can
only be adapted slowly over time, there is an implicit assumption
in (\ref{eq:GNGD}) that $E\left\{ v^{2}\left(n\right)\right\} $
also varies slowly. While this is a reasonable assumption in some
applications, it does not hold for scenarios like echo cancellation,
where the interference is speech (double-talk) that can start or stop
at any time.

\section{Proposed Algorithm For Non-Stationary Signals\label{sec:INLMS-algorithm}}

In previous work \cite{ValinAEC2006}, we proposed to use (\ref{eq:Residual-to-output-rate})
directly to adapt the learning rate. While $E\left\{ e^{2}\left(n\right)\right\} $
can easily be estimated, the estimation of the residual echo $E\left\{ r^{2}\left(n\right)\right\} $
is difficult because one does not have access to the real filter coefficients.
One reasonable assumption we can make is that:
\begin{align}
E\left\{ r^{2}\left(n\right)\right\}  & =\eta\left(n\right)E\left\{ y^{2}\left(n\right)\right\} \nonumber \\
 & \approx\eta\left(n{-}1\right)E\left\{ \hat{y}^{2}\left(n\right)\right\} \label{eq:residual-approx}
\end{align}
where $\eta\left(n\right)$ is a form of normalised filter misalignment,
and is easier to estimate than $E\left\{ r^{2}\left(n\right)\right\} $
directly, because it is assumed to vary slowly as a function of time.
Although it is possible to estimate $\eta\left(n\right)$ directly
through linear regression, the estimation remains a difficult problem. 

In this paper we propose to apply a gradient adaptive approach using
$\eta\left(n\right)$ as the control parameter. By substituting (\ref{eq:residual-approx})
into (\ref{eq:Residual-to-output-rate}), we obtain the learning rate:
\begin{equation}
\mu\left(n\right)=\min\left(\eta\left(n{-}1\right)\frac{\widehat{\sigma_{\hat{y}}^{2}}\left(n\right)}{\widehat{\sigma_{e}^{2}}\left(n\right)},1\right)\label{eq:gradient-learning-rate}
\end{equation}
where $\widehat{\sigma_{\hat{y}}^{2}}\left(n\right)$ and $\widehat{\sigma_{e}^{2}}\left(n\right)$
are respectively the estimates for $E\left\{ \hat{y}^{2}\left(n\right)\right\} $
and $E\left\{ e^{2}\left(n\right)\right\} $ and the upper bound imposed
by the $\min\left(\cdot\right)$ reflects the fact that the optimal
learning rate can never exceed unity.

\subsection{Adaptation}

In this paper we bypass the difficulty of estimating $\eta(n)$ directly
and instead propose a closed-loop gradient adaptive estimation of
$\eta(n)$. The parameter $\eta(n)$ is no longer an estimate of the
normalised misalignment, but is instead adapted in closed-loop in
such a way as to achieve a fast convergence of the adaptive filter. 

As with other gradient-adaptive methods we update the control parameter
$\eta(n)$ by computing the derivative of the squared error $\mathcal{E}\left(n\right)=\frac{1}{2}e^{*}\left(n\right)e\left(n\right)$,
this time with respect to $\eta\left(n-1\right)$, using the chain
derivation rule without the independence assumption \cite{Benveniste,Mandic2005}:
\begin{align}
\frac{\partial\mathcal{E}\left(n\right)}{\partial\eta\left(n{-}1\right)} & =\frac{1}{2}\left(\frac{\partial e^{*}\left(n\right)}{\partial\eta\left(n{-}1\right)}e\left(n\right)+e^{*}\left(n\right)\frac{\partial e\left(n\right)}{\partial\eta\left(n{-}1\right)}\right)\nonumber \\
 & =-\Re\left\{ \frac{e\left(n\right)\mathbf{x}^{H}\left(n\right)\boldsymbol{\psi}\left(n-1\right)}{\left\Vert \mathbf{x}\left(n\right)\right\Vert ^{2}}\right\} \frac{\widehat{\sigma_{\hat{y}}^{2}}\left(n\right)}{\widehat{\sigma_{e}^{2}}\left(n\right)}\label{eq:gradient-eta}
\end{align}
where
\begin{equation}
\boldsymbol{\psi}\left(n\right)=\left[\mathbf{I}-\frac{\mu\left(n\right)\mathbf{x}\left(n\right)\mathbf{x}^{H}\left(n\right)}{\left\Vert \mathbf{x}\left(n\right)\right\Vert ^{2}}\right]\boldsymbol{\psi}\left(n{-}1\right)+\mathbf{x}\left(n\right)e^{*}\left(n\right)\label{eq:benveniste-psy}
\end{equation}
is a smoothed version of the gradient. We further rewrite the update
of $\boldsymbol{\psi}\left(n\right)$ in (\ref{eq:benveniste-psy})
as
\begin{align}
\boldsymbol{\psi}\left(n\right)= & \boldsymbol{\psi}\left(n{-}1\right)-\frac{\mu\left(n\right)}{\left\Vert \mathbf{x}\left(n\right)\right\Vert ^{2}}\mathbf{x}\left(n\right)\left[\mathbf{x}^{H}\left(n\right)\boldsymbol{\psi}\left(n{-}1\right)\right]\nonumber \\
 & +\mathbf{x}\left(n\right)e^{*}\left(n\right)\label{eq:benveniste-simple}
\end{align}
so that it does not require a matrix-by-vector multiplication.

Based on this derivative, we propose the following exponential update
of $\eta\left(n\right)$. We propose to use an exponential update
in place of a more standard additive update since the misalignment
has a large dynamic range; and we want the step size to scale with
the value of $\eta\left(n\right)$. The exponential update is given
as follows:
\begin{equation}
\eta\left(n\right)=\eta\left(n-1\right)\exp\left(\frac{\rho}{\widehat{\sigma_{e}^{2}}\left(n\right)}\frac{\partial\mathcal{E}\left(n\right)}{\partial\eta\left(n{-}1\right)}\right)
\end{equation}
where $\rho$ is a step size and we have normalised the gradient $\frac{\partial\mathcal{E}\left(n\right)}{\partial\eta\left(n{-}1\right)}$
by $\widehat{\sigma_{e}^{2}}\left(n\right)$ to obtain a non-dimensional
value.

It remains to estimate $\widehat{\sigma_{\hat{y}}^{2}}\left(n\right)$
and $\widehat{\sigma_{e}^{2}}\left(n\right)$. For $\widehat{\sigma_{e}^{2}}\left(n\right)$
we have the following recursive estimator with time constant $N$
($\widehat{\sigma_{\hat{y}}^{2}}\left(n\right)$ is estimated similarly):
\begin{equation}
\hat{E}_{N}\left\{ \left|e\left(n\right)\right|^{2}\right\} =\left(1{-}\frac{1}{N}\right)\hat{E}_{N}\left\{ \left|e\left(n-1\right)\right|^{2}\right\} +\frac{1}{N}\left|e\left(n\right)\right|^{2}
\end{equation}
The question then becomes what value of $N$ to use. To maximise stability,
a conservative approach is to err on the side of picking the smallest
$\widehat{\sigma_{\hat{y}}^{2}}\left(n\right)$ and the biggest $\widehat{\sigma_{e}^{2}}\left(n\right)$
out of the set of estimated obtained by varying $N$. For efficiency,
we have chosen a subset of all possible $N$ values. The values $N=3$
and $N=10$ provide good short- to medium-term term estimation, though
the algorithm is not very sensitive to the exact choice of $N$. For
the estimation of $\widehat{\sigma_{e}^{2}}\left(n\right)$, we also
include $N=1$ to make sure that even an instantaneous onset of interference
cannot cause the filter to diverge. The complete algorithm is summarised
in Fig. \ref{fig:INLMS-summary}. 

The last aspect that needs to be addressed is the initial condition.
When the filter is initialised, all the weights are set to zero ($\hat{\mathbf{h}}\left(0\right)=\mathbf{0}$),
which means that $\hat{y}\left(n\right)=0$ and no adaptation can
take place in (\ref{eq:gradient-learning-rate}) and (\ref{eq:gradient-eta}).
In order to bootstrap the adaptation process, the learning rate $\mu\left(n\right)$
is set to a fixed constant (we use $\mu=0.25$) for a short time (until
(\ref{eq:gradient-learning-rate}) gives $\mu\left(n\right)>0.1$).
This \emph{ad hoc} procedure is only necessary when the filter is
initialised and is not required in case of echo path change. In practice,
any method that provides a small initial convergence can be used.

\begin{figure}
\begin{center}\fbox{$\begin{array}{rl}
\hat{y}\left(n\right)= & \!\!\!\!=\hat{\mathbf{h}}\left(n{-}1\right)\mathbf{x}\left(n\right)\\
e\left(n\right)= & \!\!\!\!d\left(n\right)-\hat{y}\left(n\right)\\
\widehat{\sigma_{\hat{y}}^{2}}\left(n\right)= & \!\!\!\!\min\left(\hat{E}_{3}\left\{ \left|\hat{y}\left(n\right)\right|^{2}\right\} ,\hat{E}_{10}\left\{ \left|\hat{y}\left(n\right)\right|^{2}\right\} \right)\\
\widehat{\sigma_{e}^{2}}\left(n\right)= & \!\!\!\!\max\!\left(\!\hat{E}_{1}\!\left\{ \!\left|e\left(n\right)\right|^{2}\!\right\} \!,\hat{E}_{3}\!\left\{ \!\left|e\left(n\right)\right|^{2}\!\right\} \!,\hat{E}_{10}\!\left\{ \!\left|e\left(n\right)\right|^{2}\!\right\} \!\right)\!\!\\
\mu\left(n\right)= & \!\!\!\!\min\left(\nu\left(n{-}1\right)\frac{\widehat{\sigma_{\hat{y}}^{2}}\left(n\right)}{\widehat{\sigma_{e}^{2}}\left(n\right)},1\right)\\
\hat{\mathbf{h}}\left(n\right)= & \!\!\!\!\hat{\mathbf{h}}\left(n{-}1\right)+\frac{\mu\left(n\right)}{\left\Vert \mathbf{x}\left(n\right)\right\Vert ^{2}}e^{*}\left(n\right)\mathbf{x}\left(n\right)\\
\eta\left(n\right)= & \!\!\!\!\eta\left(n{-}1\right)\exp\!\left[\frac{\rho\widehat{\sigma_{\hat{y}}^{2}}\left(n\right)\Re\left\{ e\left(n\right)\mathbf{x}^{H}\left(n\right)\boldsymbol{\psi}\left(n-1\right)\right\} }{\widehat{\sigma_{e}^{2}}\left(n\right)\left\Vert \mathbf{x}\left(n\right)\right\Vert ^{2}\widehat{\sigma_{e}^{2}}\left(n\right)}\right]\\
\boldsymbol{\psi}\left(n\right)= & \!\!\!\!\boldsymbol{\psi}\left(n{-}1\right)-\frac{\mu\left(n\right)}{\left\Vert \mathbf{x}\left(n\right)\right\Vert ^{2}}\mathbf{x}\left(n\right)\left[\mathbf{x}^{H}\left(n\right)\boldsymbol{\psi}\left(n{-}1\right)\right]\\
 & \!\!\!\!+e^{*}\left(n\right)\mathbf{x}\left(n\right)
\end{array}$}\end{center}

\caption{Summary of the INLMS algorithm\label{fig:INLMS-summary}}
\end{figure}

\subsection{Analysis}

The adaptive learning rate described above is able to deal with both
double-talk and echo path change without explicit modelling. From
(\ref{eq:gradient-learning-rate}), we can see that when the interference
changes abruptly, the denominator $\widehat{\sigma_{e}^{2}}\left(n\right)$
rapidly increases, causing an instantaneous decrease in the learning
rate. In the case of a stationary interference, the learning rate
depends on both the presence of an input signal and on the misalignment
estimate. As the filter misalignment becomes smaller, the learning
rate also becomes smaller. When the echo path changes, the gradient
starts pointing steadily in the same direction, thus significantly
increasing $\eta(n)$, which is a clear sign that the filter is no
longer properly adapted.

In gradient adaptive methods \cite{Benveniste,Mathews1993,Ang2001},
the implicit assumption is that both the near-end and the far-end
signals are nearly stationary. We have shown that the GNGD algorithm
\cite{Mandic2004} only requires the near-end signal to be nearly
stationary. In the proposed INLMS method, both signals can be non-stationary,
which is a requirement for double-talk robustness. 

It should be noted that the per-sample complexity of the proposed
algorithm only differs from the complexity of the ``classic'' algorithm
in \cite{Benveniste} by a constant ($O(1)$). For example, the total
increase in complexity for the real-valued case is due to the increased
cost of computing $\eta\left(n\right)$ and amounts to only 23 multiplications,
5 additions, 2 divisions and 1 exponential. Considering that the algorithms
have an $O(L)$ complexity ($L$ is the filter length), the difference
is negligible for any reasonable filter length.

\section{Results And Discussion\label{sec:Results-And-Discussion}}

We compare three algorithms:
\begin{itemize}
\item \textbf{Direct} learning rate adaptation \cite{Benveniste}
\item Generalized normalized gradient descent (\textbf{GNGD}) \cite{Mandic2004}
\item \textbf{INLMS} algorithm (proposed)
\end{itemize}
In each case, we use a 32-second test sequence sampled at 8 kHz with
an abrupt change in the unknown system $\mathbf{h}\left(n\right)$
at 16 seconds. The impulse responses are taken from ITU-T recommendation
G.168 (impulse responses D.7 and D.9) and the filter length is 128
samples (16 ms). We choose $\rho=0.005$ since it gave good results
over a wide range of operating conditions ($\rho$ should be inversely
proportional to the filter length). To make the comparison fair, we
also used the exponential update for the Direct method ($\rho=0.0005$
gave the best results) and the GNGD method ($\rho=0.005$ gave the
best results).

\begin{figure}
\begin{center}\includegraphics[width=75mm]{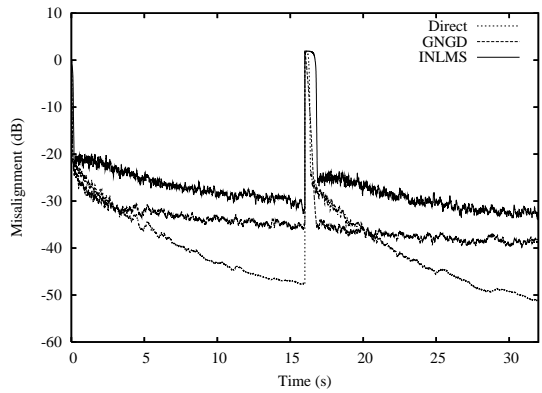}\end{center}

\caption{Normalised misalignment for white Gaussian input and interference
(scenario 1) with an abrupt change in the unknown system $\mathbf{h}\left(n\right)$
at 16 seconds. All algorithms converge.\label{fig:Scenario1}}
\end{figure}

\begin{figure}
\begin{center}\includegraphics[width=75mm]{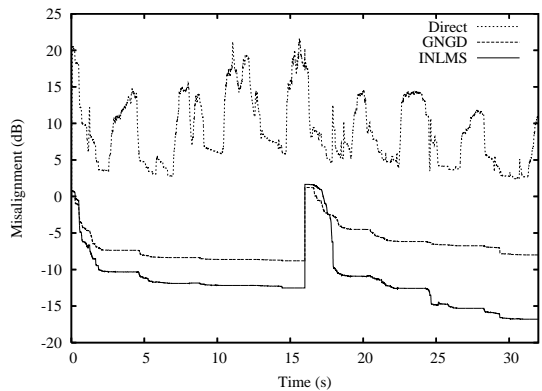}\end{center}

\caption{Normalised misalignment for typical speech input and white Gaussian
interference (scenario 2) with an abrupt change in the unknown system
$\mathbf{h}\left(n\right)$ at 16 seconds. The direct algorithm diverges.\label{fig:Scenario2}}
\end{figure}

\begin{figure}
\begin{center}\includegraphics[width=75mm]{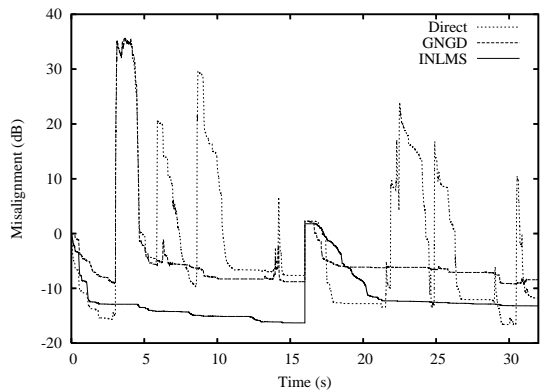}\end{center}

\caption{Normalised misalignment for typical speech input and interference
(scenario 3) with an abrupt change in the unknown system $\mathbf{h}\left(n\right)$
at 16 seconds. Note that the direct method frequently diverges, and
the GNGD method diverges less often, but still significantly between
3-4 seconds, at 6 seconds, 14 seconds and 31 seconds. All the divergence
events are due to double-talk, except at 3-4 seconds, where only interference
is present. \label{fig:Scenario3}}
\end{figure}

We test the algorithms for three scenarios:
\begin{enumerate}
\item Both the input $x\left(n\right)$ and the interference $v\left(n\right)$
are white Gaussian noise (Fig. \ref{fig:Scenario1})
\item The input $x\left(n\right)$ is speech and the interference $v\left(n\right)$
is white Gaussian noise (Fig. \ref{fig:Scenario2})
\item Both the input $x\left(n\right)$ and interference $v\left(n\right)$
are speech (Fig. \ref{fig:Scenario3}) with frequent overlap (double-talk)
\end{enumerate}
The normalised misalignment is defined as:
\begin{equation}
\Lambda\left(n\right)=\left\Vert \hat{\mathbf{h}}\left(n\right)-\mathbf{h}\left(n\right)\right\Vert ^{2}/\left\Vert \mathbf{h}\left(n\right)\right\Vert ^{2}\label{eq:misalignment}
\end{equation}

In Fig. \ref{fig:Scenario1} we can see that all three algorithms
successfully converge, albeit with differing convergence rates. In
Fig. \ref{fig:Scenario2} it can be observed that the direct algorithm
fails to converge for scenario 2 where the input is non-stationary,
but GNGD and INLMS perform well. Finally in Fig. \ref{fig:Scenario3}
we see that when the interference is non-stationary, only the proposed
INLMS algorithm performs well.

\section{Conclusion\label{sec:Conclusion}}

We have proposed a new interference-normalised least mean square (INLMS)
algorithm, based on the gradient-adaptive learning rate class of algorithms.
We have demonstrated that unlike other gradient-adaptive methods,
it is robust to non-stationarity of both the input and interference
signals. This robustness is achieved by using a control parameter
whose optimal value is independent of the power of the input and interference
signals and instead depends only on the filter misalignment. This
allows the instantaneous learning rate to react very quickly even
though the control parameter cannot.

\bibliographystyle{IEEEtran}
\bibliography{echo}

\end{document}